# High-pressure effects on superconducting properties and crystal structure of Bi-based layered superconductor La$_2$O$_2$Bi$_3$Ag$_{0.6}$Sn$_{0.4}$S$_6$

Supeng Liu[1], Ryo Matsumoto[2], Rajveer Jha[1,3], Aichi Yamashita[1], Saori I. Kawaguchi[4], Yosuke Goto[1], Yoshihiko Takano[3], Yoshikazu Mizuguchi[1]*

1. Department of Physics, Tokyo Metropolitan University, 1-1, Minami-osawa, Hachioji, 192-0397, Japan.
2. International Center for Young Scientists (ICYS), National Institute for Materials Science (NIMS), 1-1, Sengen, Tsukuba, 305-0047, Japan.
3. International Center for Materials Nanoarchitectonics (MANA), National Institute for Materials Science (NIMS), 1-1, Sengen, Tsukuba, 305-0047, Japan.
4. Japan Synchrotron Radiation Research Institute, SPring-8, 1-1-1 Kouto, Sayo, Hyogo 679-5198, Japan

* Corresponding author: Yoshikazu Mizuguchi (mizugu@tmu.ac.jp)

**Abstract**

The effects of pressure on the superconducting properties of a Bi-based layered superconductor La$_2$O$_2$Bi$_3$Ag$_{0.6}$Sn$_{0.4}$S$_6$, which possesses a four-layer-type conducting layer, have been studied through the electrical resistance and magnetic susceptibility measurements. The crystal structure under pressure was examined using synchrotron X-ray diffraction at SPring-8. In the low-pressure regime, bulk superconductivity with a transition temperature $T_c$ of ~ 4.5 K was induced by pressure, which was achieved by in-plane chemical pressure effect owing to the compression of the tetragonal structure. In the high-pressure regime above 6.4 GPa, a structural symmetry lowering was observed, and superconducting transitions with a $T_c$ ~ 8 K were observed. Our results suggest the possible commonality on the factor essential for $T_c$ in Bi-based superconductors with two-layer-type and four-layer-type conducting layers.

## 1. Introduction

Since the discovery of BiS$_2$-based layered superconductors Bi$_4$O$_4$S$_3$ and La(O,F)BiS$_2$ in 2012 [1,2], many superconductors having superconducting BiS$_2$ layers have been synthesized [3]. The typical BiS$_2$-based superconductor system is RE(O,F)BiS$_2$ (RE: La, Ce, Pr, Nd, Sm, Yb, Bi) [2–9]. The crystal structure of RE(O,F)BiS$_2$ is composed of alternate stacks of a superconducting BiS$_2$ bilayer (Bi$_2$S$_4$ layer) and an insulating RE$_2$O$_2$ layer. In RE(O,F)BiS$_2$, electron carriers generated through the partial substitution of F for the O site are needed for the emergence of superconductivity because the pristine phase, REOBiS$_2$ with no carrier doping, is a semiconductor with a band gap [10,11]. The notable feature of La(O,F)BiS$_2$ is the superconducting characteristics like a filamentary superconductor, in which the superconducting volume fraction is significantly low as compared to a bulk superconductor [2]. To induce bulk superconductivity in La(O,F)BiS$_2$, two pressure effects were used. One route to induce bulk superconductivity is application of *external pressures*. The transition temperature ($T_c$) of LaO$_{0.5}$F$_{0.5}$BiS$_2$ increases from 2 K to 10–11 K by the effect of external pressures, and the superconducting characteristics of the high-pressure phase of LaO$_{0.5}$F$_{0.5}$BiS$_2$ is bulk in nature [2,12–15]. The dramatic improvement of the superconducting properties can be explained by a structural transition from tetragonal ($P4/nmm$) to monoclinic ($P2_1/m$), which takes place at quite low pressure of ~1 GPa [15]. In addition, the structural transition largely affects electronic structure and the bonding nature in the crystal structure [15,16]. Another route to induce bulk superconductivity in the RE(O,F)BiS$_2$ system is application of *chemical pressures*. The chemical pressure effect can be generated through isovalent substitutions. For example, partial substitutions of La by smaller RE induce bulk superconductivity and increases $T_c$ up to 5-6 K in RE(O,F)BiS$_2$ with RE = Nd, Sm [9]. Furthermore, the partial substitutions of S at the conducting layer by larger Se are also effective to induce bulk superconductivity [17]. Those chemical pressure effects do not induce a structural transition but compresses tetragonal lattice. The compression of the Bi-Ch (Ch: S, Se) square network suppresses intrinsic structural disorder [18–20], which is present due to the presence of Bi lone pairs, in RE(O,F)BiS$_2$ [21].

The target phase of this study is La$_2$O$_2$Bi$_3$Ag$_{0.6}$Sn$_{0.4}$S$_6$, whose basic crystal structure can be regarded as the stacking of La$_2$O$_2$Bi$_2$S$_4$ and M$_2$S$_2$ (M = Bi$_{0.5}$Ag$_{0.3}$Sn$_{0.2}$) [22]. In La$_2$O$_2$Bi$_2$M$_2$S$_6$-type structure, a rock-salt-type M$_2$S$_2$ layer is inserted into the van-der-Waals gap of LaOBiS$_2$ [23–25]. Regarding the superconducting layer of RE(O,F)BiS$_2$ as *two-layer-type,* we can classify that of La$_2$O$_2$Bi$_2$M$_2$S$_6$ as *four-layer-type* (see Fig. 1). The first La$_2$O$_2$Bi$_2$M$_2$S$_6$-type compound was obtained with M = Pb and was reported as a thermoelectric material [23]. In 2018, we observed superconductivity in La$_2$O$_2$Bi$_3$AgS$_6$ (M = Bi$_{0.5}$Ag$_{0.5}$) below 0.5 K [26]. The low $T_c$ in La$_2$O$_2$Bi$_3$AgS$_6$ was found to be related to the charge-density-wave-like (CDW-like) transition, and the partial substitution of Sn for the Ag site was effective to suppress the transition and improve superconducting properties in La$_2$O$_2$Bi$_3$Ag$_{0.6}$Sn$_{0.4}$S$_6$, in which $T_c$ reaches 2.5 K [22]. Furthermore, application of chemical pressures in La$_2$O$_2$Bi$_3$Ag$_{0.6}$Sn$_{0.4}$S$_6$ further improved $T_c$ up to 4 K in Eu-substituted La$_{1.6}$Eu$_{0.4}$O$_2$Bi$_3$Ag$_{0.6}$Sn$_{0.4}$S$_6$ and similar phases with various RE [27–29]. On the basis of the knowledge on the dramatic external pressure effects in two-layer-type BiS$_2$-based systems, we expected further increases in $T_c$ by external pressures in La$_2$O$_2$Bi$_3$Ag$_{0.6}$Sn$_{0.4}$S$_6$. Therefore, in this study, we have investigated the superconducting properties and crystal structure of La$_2$O$_2$Bi$_3$Ag$_{0.6}$Sn$_{0.4}$S$_6$ under high pressures.

## 2. Experimental details

A polycrystalline sample of La$_2$O$_2$Bi$_3$Ag$_{0.6}$Sn$_{0.4}$S$_6$ was prepared by solid-state reaction. Powders of La$_2$S$_3$ (99.9%), Bi$_2$O$_3$ (99.999%), Ag$_2$O (99%up), Bi (99.999%), Sn (99.9%), and S (99.9999%) were mixed by mortar and pestle in air, pressed into a pellet, sealed in an evacuated quartz tube, and heated at 720 °C for 15 h. The obtained sample was ground, mixed, pelletized, and annealed under the same condition as the first sintering. The sample quality was comparable to that reported in Ref. 22, which was confirmed by laboratory powder X-ray diffraction (XRD) and energy dispersive X-ray spectroscopy (EDX).

Electrical resistance measurements were performed at ambient pressure and high pressures using a standard four-probe method on a Physical Property Measurement System (Quantum Design: PPMS). A diamond anvil cell (DAC) with boron-doped diamond electrodes was used for the high-pressure measurements [30,31]. One side of diamond anvil equipped the electrodes and the polycrystalline sample was mounted on the center part of anvil, as shown in Fig. 2. The opposite anvil is culet-type with a diameter of 400 μm. A stainless-steel (SUS316) and a mixture of cubic boron nitride and ruby powders were used as a sample chamber and a pressure-transmitting medium, respectively. The applied pressure was estimated from the relationship between the pressure and wavelength of the ruby fluorescence [32] measured by an inVia Raman

Microscope (RENISHAW). The temperature dependence of magnetic susceptibility was measured using a superconducting quantum interference devise (SQUID) magnetometer (MPMS-3, Quantum Design) under pressures from ambient pressure to ~1.2 GPa. To apply pressures, a piston cylinder cell for MPMS-3 was used. A pressure medium was Daphne 7373, and Pb was used as a manometer. A magnetic field used for the magnetization experiment was 10 Oe. The applied pressure was estimated using a $T_c$ of Pb.

The crystal structure investigation was performed using synchrotron X-ray at the beamline BL10XU of SPring-8 under the proposal (No.: 2020A2051). The powder sample was loaded into a sample hole drilled in a SUS gasket together with a pressure medium, He gas. The samples were compressed to the pressures of interest using single crystal diamond anvils with 600 and 400 μm culet. Pressure was determined by wavelength shift of the ruby R1 fluorescence line [33]. The wavelength was 0.413102 Å (Run-1) and 0.41315 Å (Run-2). The experiments Run-1 and Run-2 were performed using different powders. The obtained powder X-ray diffraction (XRD) patterns were refined by Rietveld refinement using RIETAN-FP [34]. Crystal structure images were depicted using VESTA [35]. As shown in Fig. 1(b), we regarded that the outer metal site (in the $BiS_2$ layer) is occupied by Bi only according to the site selectivity found in previous structural analyses [22,24,25]. Instead, the inner metal site (the M site of the MS layer) is regarded as the solution of Bi, Ag, and Sn. In the refinements, the isotropic displacement parameters ($B$) for the La, O, and S sites were fixed to 1.

## 3. Results and discussion
### 3-1. Electrical resistance

To investigate the pressure evolution of $T_c$, the electrical resistance was measured on DAC. Figure 3 shows the temperature dependence of the electrical resistance of $La_2O_2Bi_3Ag_{0.6}Sn_{0.4}S_6$ at ambient pressure. The $T_c^{onset}$ was ~2.5 K, which is consistent with that reported in Ref. 22. Figure 4(a) shows the temperature dependences of the resistance under pressures up to 11.2 GPa. The low-temperature resistance data near the superconducting transition is plotted in Fig. 4(b), and the temperature dependences of normalized resistance $R(T) / R(T = 8 K)$ under pressures up to 20.5 GPa are plotted in Fig. 4(c). At 4.6 GPa, a jump of $T_c^{onset}$ was seen, and the $T_c^{onset}$ reaches 8 K at 11.2 GPa. To confirm that the resistance drops observed under high pressures were a superconducting transition, the temperature dependences of the resistance under magnetic fields were measured. As shown in Fig. 5, the $T_c^{onset}$ for the data at 0.25 and 11.2 GPa clearly decreases with increasing magnetic field. Through the magnetoresistance measurements, we confirmed that the resistance drops are a superconducting transition.

### 3-2. Magnetic susceptibility

Investigation of the temperature dependence of the magnetic susceptibility is useful to obtain information about $T_c$ and bulk nature of the superconductivity under high pressures. Figure 6(a) shows the temperature dependences of the susceptibility for $La_2O_2Bi_3Ag_{0.6}Sn_{0.4}S_6$ under pressures. As reported in Ref. 22, the superconducting states at ambient pressure is not bulk in nature, and bulk superconductivity was induced by chemical pressure effects via partial S substitution by Se or partial La substitution by smaller RE [22,27–29]. With increasing pressure, $T_c$ of $La_2O_2Bi_3Ag_{0.6}Sn_{0.4}S_6$ reached 4 K, and bulk superconductivity was induced at pressures above 0.48 GPa, which was evaluated through the emergence of large shielding volume fraction of magnetic susceptibility. The estimated magnetic $T_c^{mag}$ is plotted as a function of pressure in Fig. 6(b). See supplemental materials for the estimation of $T_c^{mag}$. The $T_c$ shows a plateau behavior at $P$ = 0.48−1.22 GPa. The origin of the plateau was discussed with the structural evolution in the following section.

### 3-3. Synchrotron powder X-ray diffraction

To discuss the evolution of superconductivity under pressures from structural point of view, we have performed SXRD experiments under high pressures. Figure 7(a) shows the powder SXRD patterns (Run-2) at 2.0–9.9 GPa. The SXRD patterns for Run-1 are displayed in Fig. S2 (Supplemental Materials). The position of the peaks shifts to higher angles with increasing pressure. Figure 7(b) shows the evolution of the 200 peak (tetragonal) by pressure. Since the 200 peak does not split below 6.4 GPa, the data suggests that the tetragonal ($P4/nmm$) [22] was maintained up to 6.4 GPa. In Figs. S3 and S4 (Supplemental Materials), typical Rietveld refinement results ($P$ = 0.2 and 6.4 GPa) analyzed with the tetragonal model are presented. For all the refinements, the reliability factor $R_{wp}$ was less than 7%. The refined lattice constants are plotted as a function of pressure in Figs. 7(c) and 7(d). For both lattice constants $a$ and $c$, a plateau was observed at $P$ = 0.4–2 GPa. This behavior is very similar to the trend in the $T_c^{mag}$-$P$ plot (Fig. 6(b)).

Therefore, the $T_c^{mag}$ plateau observed in the low-pressure regime is related to the unique compression of the lattice by pressure in the low-pressure regime.

In a high-pressure regime above 6.4 GPa, the splitting of the tetragonal 200 peak was clearly observed as shown in Fig. 7(b). The peak splitting indicates lowering of the in-plane symmetry from the tetragonal (four-fold) symmetry. Due to broadened peaks under high pressures, we could not refine the high-pressure structure from the SXRD data. From the structural analogy to two-layer-type BiS$_2$-based systems, a monoclinic ($P2_1/m$) is one of the candidate space groups [15,36,37]. However, there is another structural model suggested for the high-pressure phase of La(O,F)Bi(S,Se)$_2$, in which a stacking faults are yielded due to a loss of long-range order along the stacking direction [38]. Therefore, to clarify the crystal structure of the high-pressure phase of La$_2$O$_2$Bi$_3$Ag$_{0.6}$Sn$_{0.4}$S$_6$, single crystal structural analysis and local structure analysis are needed.

In Fig. 8, typical atomic distances are plotted as a function of pressure. Herein, we briefly describe the pressure evolution of the atomic distances. First, let us discuss about the structure of the BiS$_2$ layer. The in-plane Bi-S1 distance shows a plateau at lower pressures and decreases with increasing pressure at high pressures. Since the shrinkage of in-plane Bi-S1 distance corresponds to positive in-plane chemical pressure in the tetragonal phase [18], the improved superconducting properties can be understood by the in-plane chemical pressure effects. Similar shrinkage was found in the M-S3 distance along the in-plane direction. In addition, the decrease in the Bi-S2 distance was also seen; the relationship between the Bi-S2 distance and $T_c$ was suggested in two-layer-type systems [39]. Seeing the interlayer distance between the BiS$_2$ layer and the MS layers, we find a significant shrinkage of the Bi-S3 distance. This trend indicates that the interaction between these two layers are enhanced by applying pressure, and the structural transition finally takes place. Interestingly, the M-S3 distance along the $c$-axis shows a trend to increase above 1.2 GPa with increasing pressure. These pressure evolutions of the atomic distances suggest that the structural change of La$_2$O$_2$Bi$_3$Ag$_{0.6}$Sn$_{0.4}$S$_6$ under pressures is not simple, and dramatic structural modifications at both BiS$_2$ and MS layers are induced by pressure effect even in the tetragonal structure.

### 3-4. Phase diagram

The $T_c^{onset}$s estimated through resistance measurements are plotted as a function of pressure in Fig. 9 with the data of the lattice constant $a$. In the low-pressure regime, bulk superconductivity is induced in the tetragonal phase by pressure. The states are similar to that achieved by in-plane chemical pressure effects owing to the compression of the tetragonal structure, as described in the section 3-3. In the high-pressure regime above 6.4 GPa, a possible structural transition from tetragonal to a lower-symmetric phase was found, and the $T_c$ reaches 8 K at 11.2 GPa. The jump behavior in the $T_c$-$P$ is quite similar to that observed in RE(O,F)BiS$_2$ [13-15] and related BiS$_2$-based systems such as EuFBiS$_2$ and (Sr,La)FBiS$_2$, which show a structural transition from tetragonal to monoclinic [36,37]. With increasing pressure above 12 GPa, $T_c$ tends to slightly decrease. The high-pressure behavior also resembles that observed in LaO$_{0.5}$F$_{0.5}$BiS$_2$ [15]. As reported in Refs. 24 and 40, the band structure, particularly band dispersion near the Fermi Energy, is clearly different between two-layer-type BiS$_2$-based systems like RE(O,F)BiS$_2$ and four-layer-type La$_2$O$_2$Bi$_2$M$_2$S$_6$. However, the observed $T_c$-$P$ phase diagrams exhibit clear similarity: the jump in $T_c$ is observed where a structural transition is observed. Although we need to further investigate the high-pressure structure for La$_2$O$_2$Bi$_3$Ag$_{0.6}$Sn$_{0.4}$S$_6$ and other La$_2$O$_2$Bi$_2$M$_2$S$_6$-type materials to conclude on the commonality between two-layer-type and four-layer-type systems, the present results imply that there is a commonality on the factor essential for $T_c$ in Bi-based layered superconductors with the two-layer-type and four-layer-type conducting layers.

## 4. Conclusion

In this study, we have investigated the effects of pressure on the superconducting properties of a polycrystalline sample of La$_2$O$_2$Bi$_3$Ag$_{0.6}$Sn$_{0.4}$S$_6$ through the electrical resistance measurements on a DAC and the magnetic susceptibility measurements on a piston cylinder cell. The crystal structure under pressure was examined using SXRD and Rietveld refinement for the tetragonal phase. In the low-pressure regime below 2.4 GPa, bulk superconductivity with a transition temperature $T_c$ of ~ 4.5 K was induced by pressure, which was achieved by compression of the tetragonal structure. In the high-pressure regime above 6.4 GPa, a structural symmetry lowering was observed. In the high-pressure regime, superconducting transitions with a $T_c$ ~ 8 K were observed. Through structural refinements, we found that the in-plane chemical pressure is essential for the improvement of the superconducting properties in the tetragonal phase. Furthermore, even in the tetragonal phase, both BiS$_2$ and MS layers show dramatic structural change under pressures. The jump of $T_c$ accompanied with a structural phase transition by applying pressure is similar to that observed in two-layer-type $BiS_2$-based systems. Therefore, there should be a commonality on the factor essential for $T_c$ in Bi-based (two-layer-type and four-layer-type) superconductors.

**Acknowledgment**

The authors thank O. Miura for his supports in magnetic susceptibility measurements. This work was partly supported by JSPS KAKENHI (Grant Nos: 18KK0076, 19H02177, and 20K22420), JST-CREST (Grant No.: JPMJCR16Q6), JST-Mirai Program (Grant No: JPMJMI17A2), and Tokyo Metropolitan Government Advanced Research (Grant Number: H31-1).

**References**

1. Y. Mizuguchi, H. Fujihisa, Y. Gotoh, K. Suzuki, H. Usui, K. Kuroki, S. Demura, Y. Takano, H. Izawa, and O. Miura, Phys. Rev. B 86, 220510 (2012).
2. Y. Mizuguchi, S. Demura, K. Deguchi, Y. Takano, H. Fujihisa, Y. Gotoh, H. Izawa, and O. Miura, J. Phys. Soc. Jpn. 81, 114725 (2012).
3. Y. Mizuguchi, J. Phys. Soc. Jpn.88, 041001 (2019).
4. J. Xing, S. Li, X. Ding, H. Yang, and H. H. Wen, Phys. Rev. B 86, 214518 (2012).
5. S. Demura, K. Deguchi, Y. Mizuguchi, K. Sato, R. Honjyo, A. Yamashita, T. Yamaki, H. Hara, T. Watanabe, S. J. Denholme, M. Fujioka, H. Okazaki, T. Ozaki, O. Miura, T. Yamaguchi, H. Takeya, and Y. Takano, J. Phys. Soc. Jpn. 84, 024709 (2015).
6. R. Jha, A. Kumar, S. K. Singh, and V. P. S. Awana, J. Supercond. Novel Magn. 26, 499 (2013).
7. S. Demura, Y. Mizuguchi, K. Deguchi, H. Okazaki, H. Hara, T. Watanabe, S. J. Denholme, M. Fujioka, T. Ozaki, H. Fujihisa, Y. Gotoh, O. Miura, T. Yamaguchi, H. Takeya, and Y. Takano, J. Phys. Soc. Jpn. 82, 033708 (2013).
8. D. Yazici, K. Huang, B. D. White, A. H. Chang, A. J. Friedman, and M. B. Maple, Philos. Mag. 93, 673 (2013).
9. J. Kajitani, T. Hiroi, A. Omachi, O. Miura, and Y. Mizuguchi, J. Phys. Soc. Jpn. 84, 044712 (2015).
10. H. Usui, K. Suzuki, and K. Kuroki, Phys. Rev. B 86, 220501 (2012).

11. K. Suzuki, H. Usui, K. Kuroki, T. Nomoto, K. Hattori, and H. Ikeda, J. Phys. Soc. Jpn. 88, 041008 (2019).

12. Y. Mizuguchi, T. Hiroi, J. Kajitani, H. Takatsu, H. Kadowaki, and O. Miura, J. Phys. Soc. Jpn. 83, 053704 (2014).

13. H. Kotegawa, Y. Tomita, H. Tou, H. Izawa, Y. Mizuguchi, O. Miura, S. Demura, K. Deguchi, and Y. Takano, J. Phys. Soc. Jpn. 81, 103702 (2012).

14. C. T. Wolowiec, D. Yazici, B. D. White, K. Huang, and M. B. Maple, Phys. Rev. B 88, 064503 (2013).

15. T. Tomita, M. Ebata, H. Soeda, H. Takahashi, H. Fujihisa, Y. Gotoh, Y. Mizuguchi, H. Izawa, O. Miura, S. Demura, K. Deguchi, and Y. Takano, J. Phys. Soc. Jpn. 83, 063704 (2014).

16. M. Ochi, R. Akashi, and K. Kuroki, J. Phys. Soc. Jpn. 85, 094705 (2016).

17. T. Hiroi, J. Kajitani, A. Omachi, O. Miura, and Y. Mizuguchi, J. Phys. Soc. Jpn. 84, 024723 (2015).

18. Y. Mizuguchi, A. Miura, J. Kajitani, T. Hiroi, O. Miura, K. Tadanaga, N. Kumada, E. Magome, C. Moriyoshi, and Y. Kuroiwa, Sci. Rep. 5, 14968 (2015).

19. Y. Mizuguchi, K. Hoshi, Y. Goto, A. Miura, K. Tadanaga, C. Moriyoshi, and Y. Kuroiwa, J. Phys. Soc. Jpn. 87, 023704 (2018).

20. E. Paris, B. Joseph, A. Iadecola, T. Sugimoto, L. Olivi, S. Demura, Y. Mizuguchi, Y. Takano, T. Mizokawa, and N. L. Saini, J. Phys.: Condens. Matter 26, 435701 (2014).

21. Y. Mizuguchi, E. Paris, T. Sugimoto, A. Iadecola, J. Kajitani, O. Miura, T. Mizokawa, and N. L. Saini, Phys. Chem. Chem. Phys. 17, 22090 (2015).

22. R. Jha, Y. Goto, T. D. Matsuda, Y. Aoki, M. Nagao, I. Tanaka, and Y. Mizuguchi, Sci. Rep. 9, 13346 (2019).

23. Y. L. Sun, A. Ablimit, H. F. Zhai, J. K. Bao, Z. T. Tang, X. B. Wang, N. L. Wang, C. M. Feng, and G. H. Cao, Inorg. Chem. 53, 11125 (2014).

24. Y. Mizuguchi, Y. Hijikata, T. Abe, C. Moriyoshi, Y. Kuroiwa, Y. Goto, A. Miura, S. Lee, S. Torii, T. Kamiyama, C. H. Lee, M. Ochi, and K. Kuroki, EPL 119, 26002 (2017).

25. Y. Hijikata, T. Abe, C. Moriyoshi, Y. Kuroiwa, Y. Goto, A. Miura, K. Tadanaga, Y. Wang, O. Miura, and Y. Mizuguchi, J. Phys. Soc. Jpn. 86, 124802 (2017).

26. R. Jha, Y. Goto, R. Higashinaka, T. D. Matsuda, Y. Aoki, Y. Mizuguchi, J. Phys. Soc. Jpn. 87, 083704 (2018).

27. R. Jha and Y. Mizuguchi, Condens. Matter 5, 27 (2020).
28. R. Jha, Y. Goto, R. Higashinaka, A. Miura, C. Moriyoshi, Y. Kuroiwa, and Y. Mizuguchi, Physica C 576, 1353731 (2020).
29. G. C. Kim, M. Cheon, W. Choi, D. Ahmad, Y. S. Kwon, R. Ko, and Y. C. Kim, J. Supercond. Nov. Magn. 33, 625 (2020).
30. R. Matsumoto, Y. Sasama, M. Fujioka., T. Irifune, M. Tanaka, T. Yamaguchi, H. Takeya, and Y. Takano, Rev. Sci. Instrum. 87, 076103 (2016).
31. R. Matsumoto, A. Yamashita, H. Hara, T. Irifune, S. Adachi, H. Takeya, and Y. Takano, Appl. Phys. Express 11, 053101 (2018).
32. G. J. Piermarini, S. Block, J. D. Barnett, and R. A. Forman, J. Appl. Phys. 46, 2774 (1975).
33. C. S. Zha, H. K. Mao, R. J. Hemley, Proc. NAtl. Acad. Sci 97, 13494 (2000).
34. F. Izumi and K. Momma, Solid State Phenom. 130, 15 (2007).
35. K. Momma and F. Izumi, J. Appl. Crystallogr. 41, 653 (2008).
36. C. Y. Guo, Y. Chen, M. Smidman, S. A. Chen, W. B. Jiang, H. F. Zhai, Y. F. Wang, G. H. Cao, J. M. Chen, X. Lu, and H. Q. Yuan, Phys. Rev. B 91, 214512 (2015).
37. A. Yamashita, R. Jha, Y. Goto, A. Miura, C. Moriyoshi, Y. Kuroiwa, C. Kawashima, K. Ishida, H. Takahashi, and Y. Mizuguchi, Sci. Rep. 10, 12880 (2020).
38. V. Svitlyk, A. Krzton-Maziopa, and M. Mezouar, Phys. Rev. B 100, 144107 (2019).
39. G. M. Pugliese, E. Paris, F. G. Capone, F. Stramaglia, T. Wakita, K. Terashima, T. Yokoya, T. Mizokawa, Y. Mizuguchi, and N. L. Saini, Phys. Chem. Chem. Phys. 22, 22217-22225 (2020)
40. K. Kurematsu, M. Ochi, H. Usui, K. Kuroki, J. Phys. Soc. Jpn. 89, 024702 (2020).

Figures

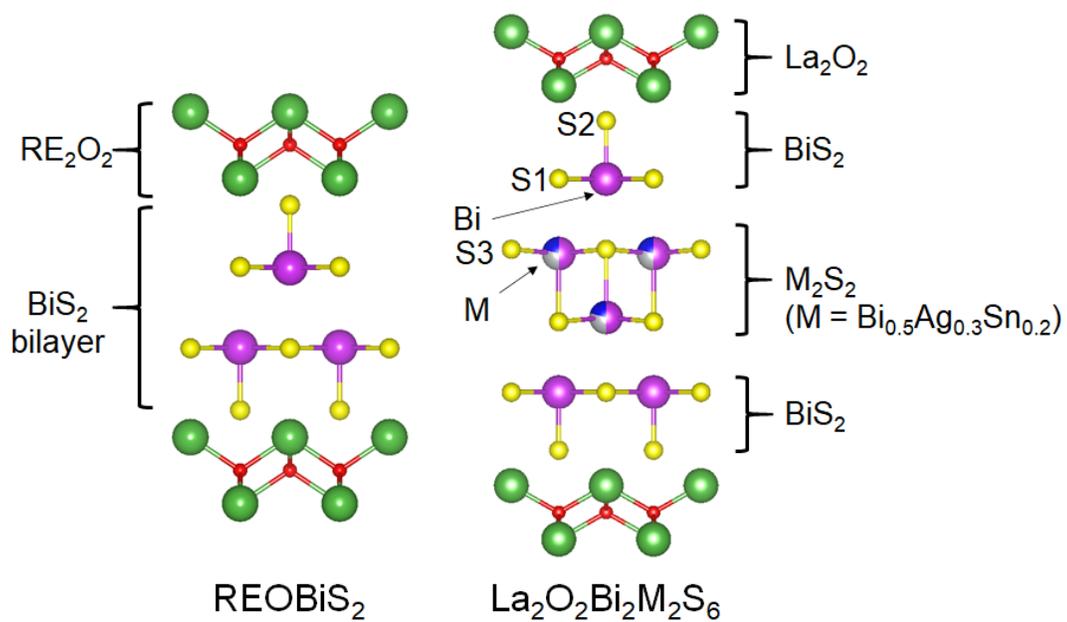

Fig. 1. Schematic images of crystal structure of REOBiS$_2$ and La$_2$O$_2$Bi$_2$M$_2$S$_6$ with M = Bi$_{0.5}$Ag$_{0.3}$Sn$_{0.2}$.

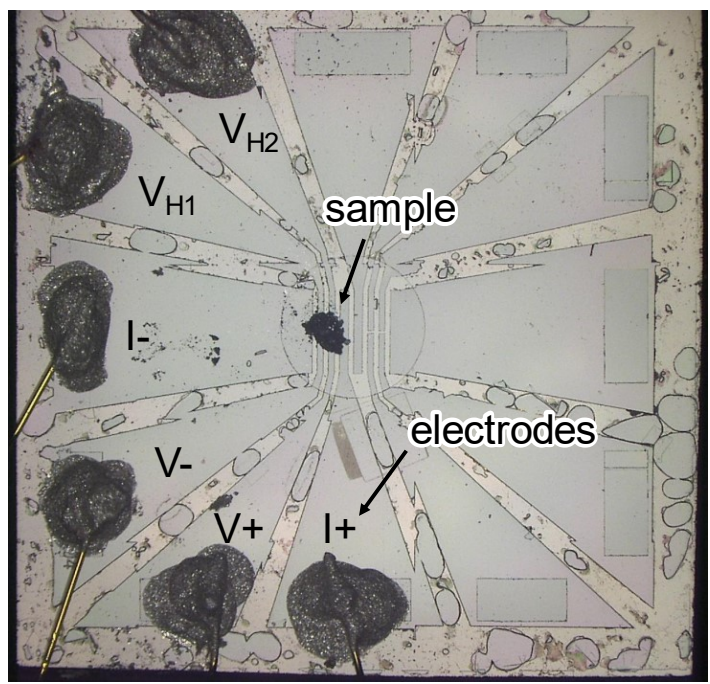

Figure 2. Optical image of the diamond anvil with boron-doped diamond electrodes.

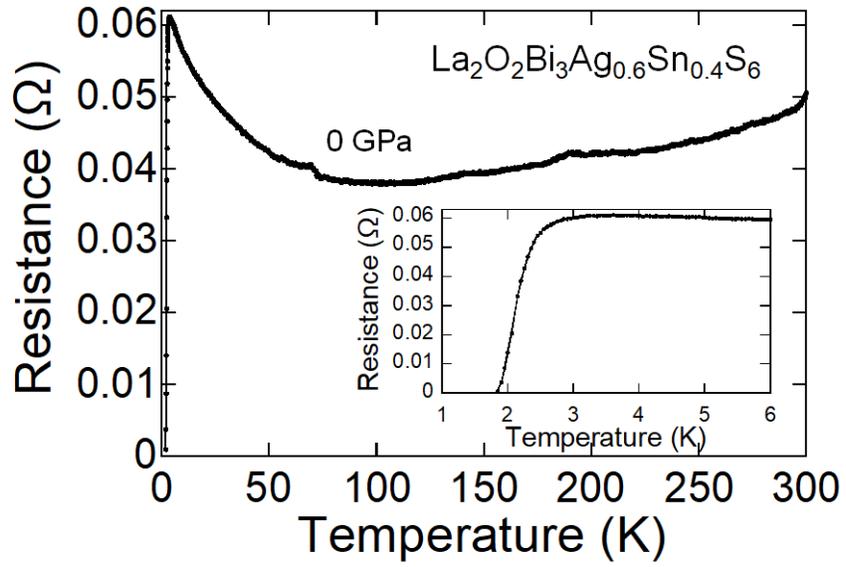

**Figure 3.** Temperature dependence of electrical resistance for $La_2O_2Bi_3Ag_{0.6}Sn_{0.4}S_6$. The inset shows low-temperature data near the superconducting transition.

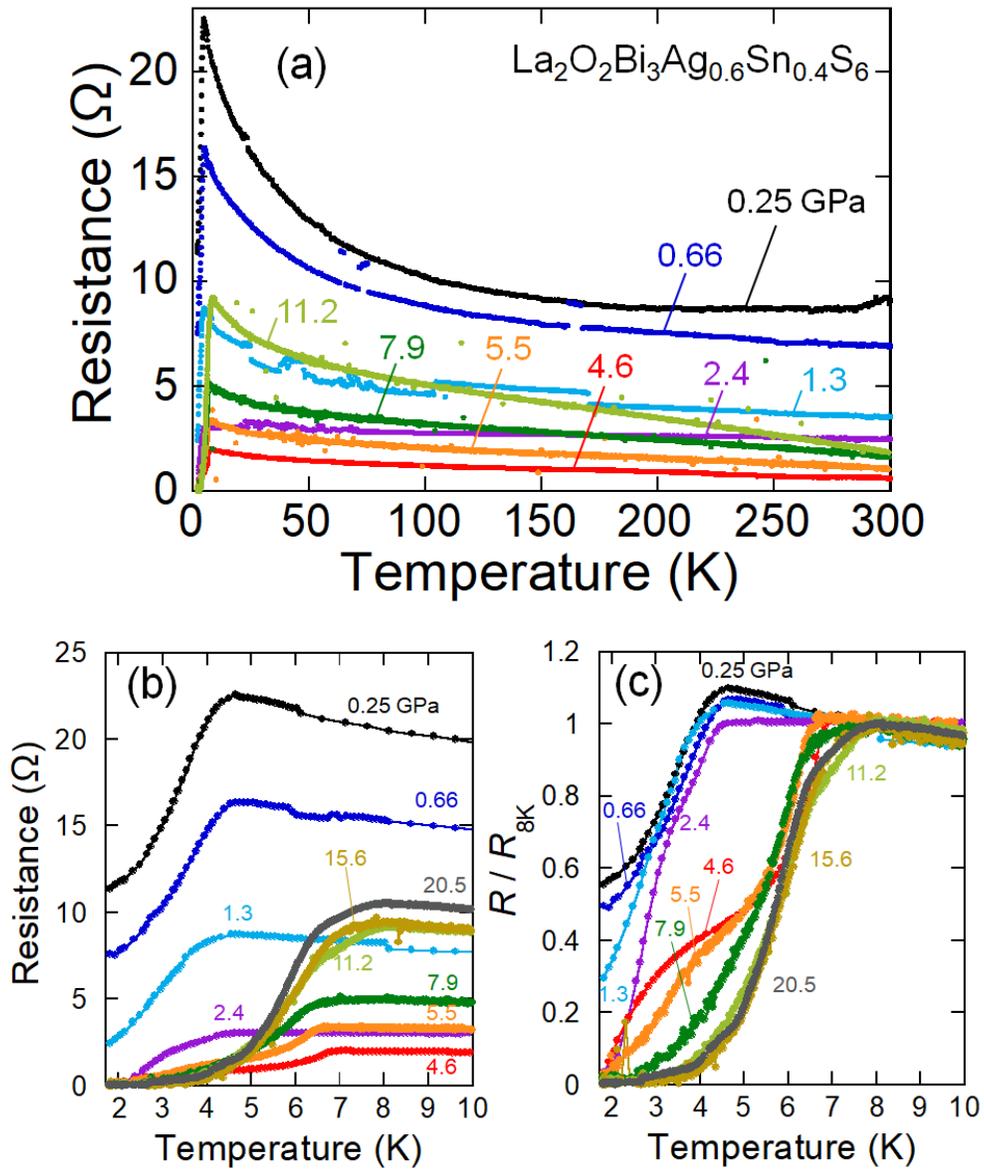

Figure 4. (a) Temperature dependences of electrical resistance for $La_2O_2Bi_3Ag_{0.6}Sn_{0.4}S_6$ under pressures. (b) Low-temperature plots of the temperature dependences of resistance. (c) Temperature dependences of normalized resistance $R(T)/R(T = 8K)$.

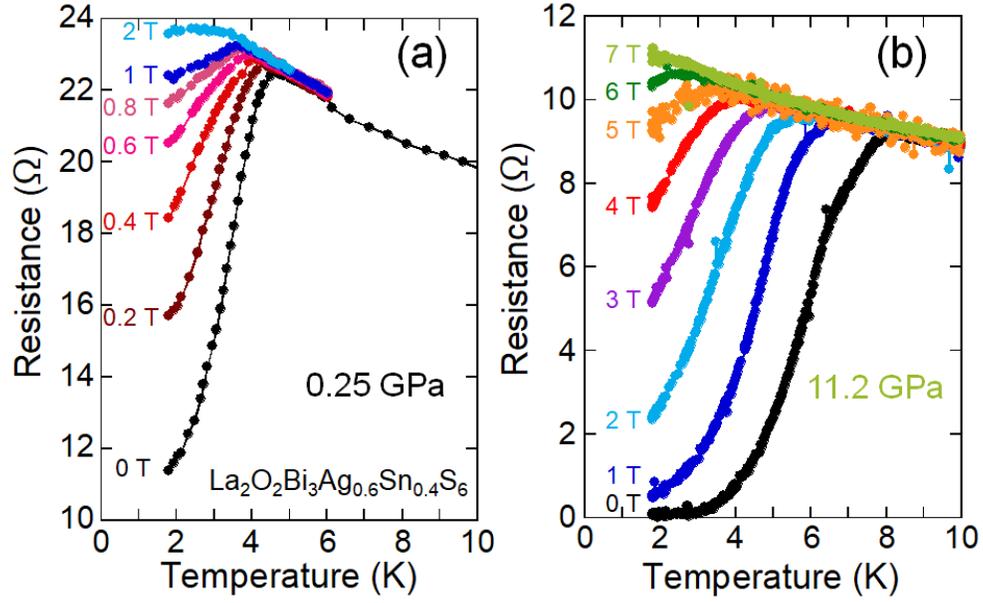

**Figure 5.** Temperature dependences of resistance at (a) 0.25 GPa and (b) 11.2 GPa under magnetic fields.

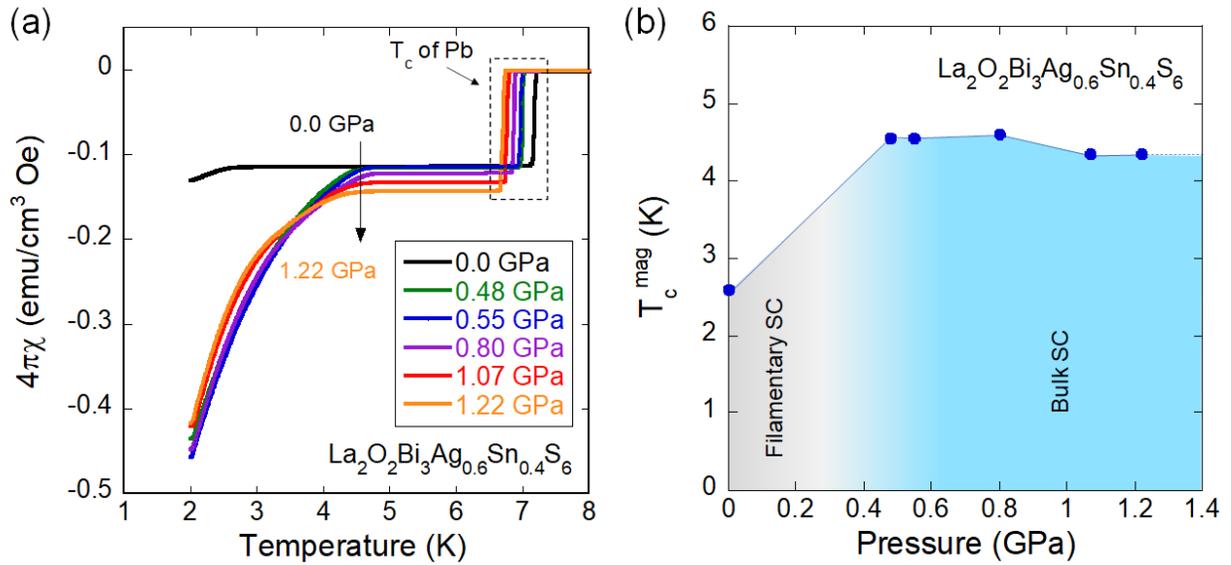

**Figure 6.** (a) Temperature dependences of magnetic susceptibility for $La_2O_2Bi_3Ag_{0.6}Sn_{0.4}S_6$ under pressures up to 1.22 GPa. (b) Pressure dependence of $T_c^{mag}$ for $La_2O_2Bi_3Ag_{0.6}Sn_{0.4}S_6$.

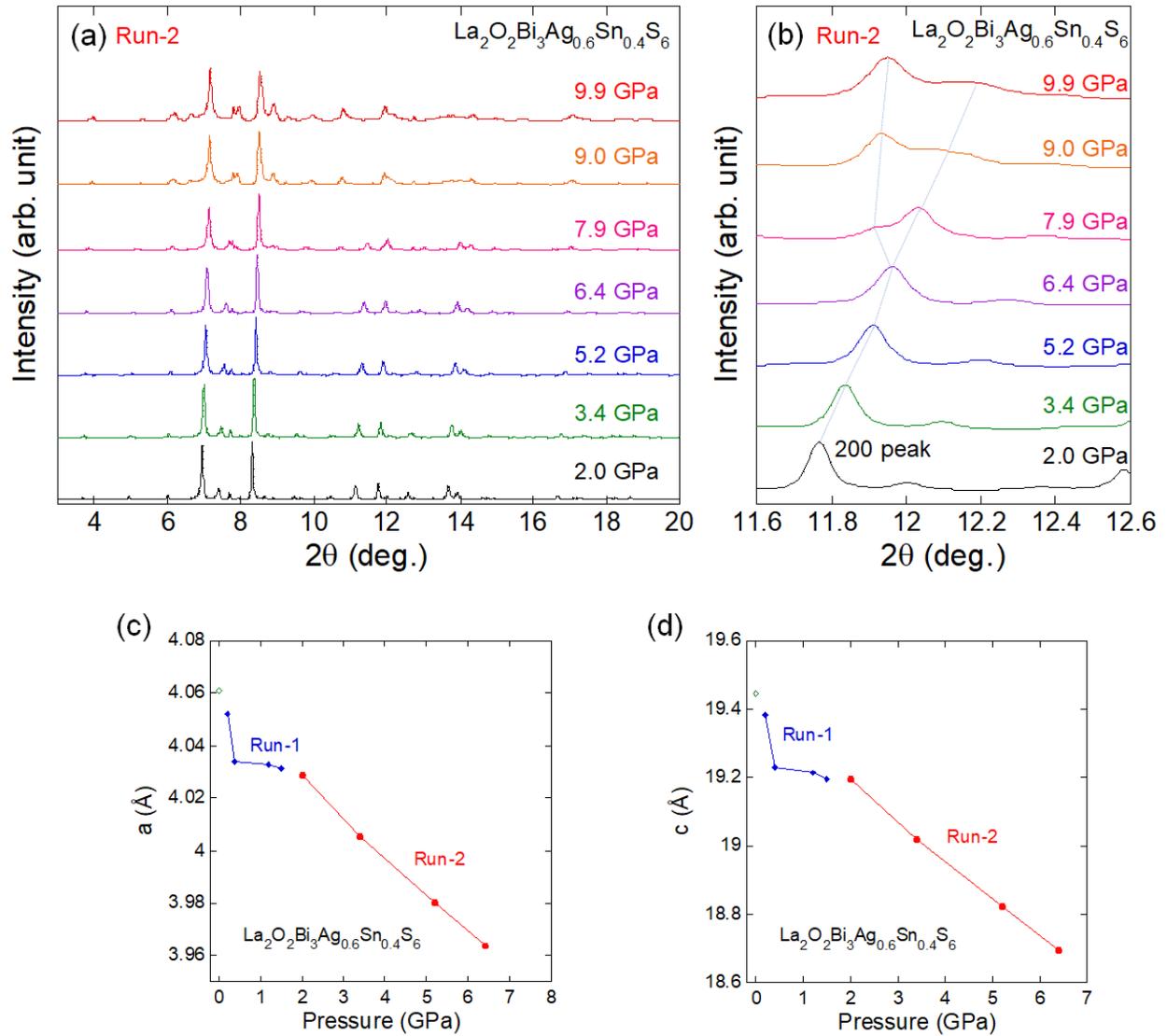

**Figure 7. (a) Powder SXRD patterns (Run-2) for $La_2O_2Bi_3Ag_{0.6}Sn_{0.4}S_6$. (b) Evolution of the 200 (tetragonal) peak under high pressure. The dashed lines are eye-guide and showing the peak splitting at a pressure between 6.4 and 7.9 GPa. (c,d) Pressure dependence of the lattice constant *a* and *c* obtained through Run-1 and Run-2.**

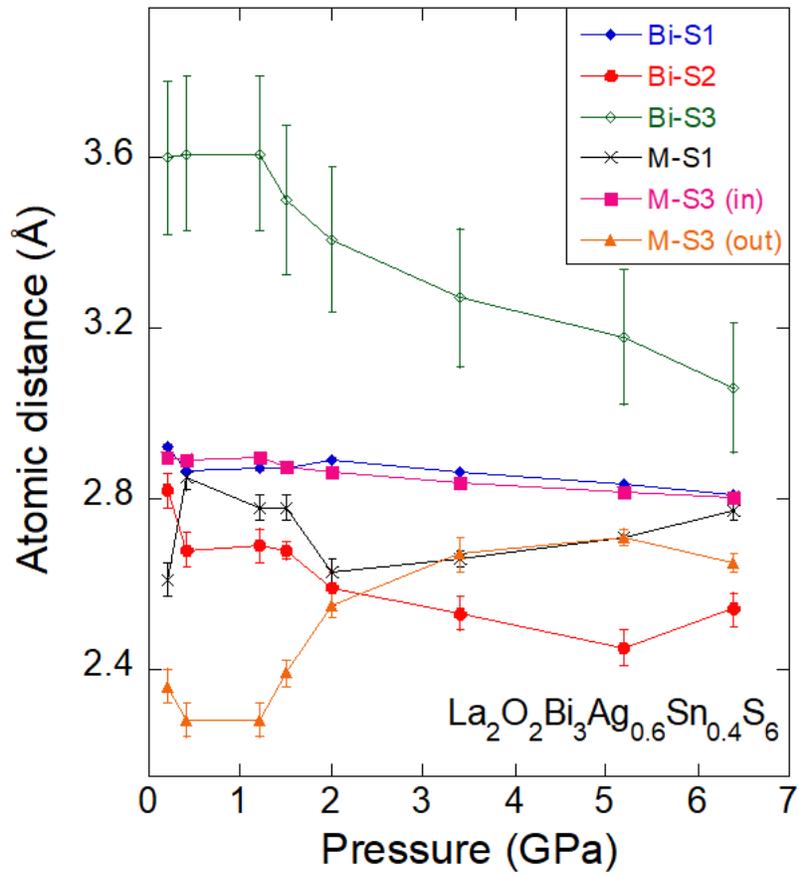

**Figure 8.** Atomic distances of La$_2$O$_2$Bi$_3$Ag$_{0.6}$Sn$_{0.4}$S$_6$ as a function of pressure. M-S3 (in) and M-S3 (out) denote the atomic distances along the in-plane direction and the *c*-axis direction, respectively.

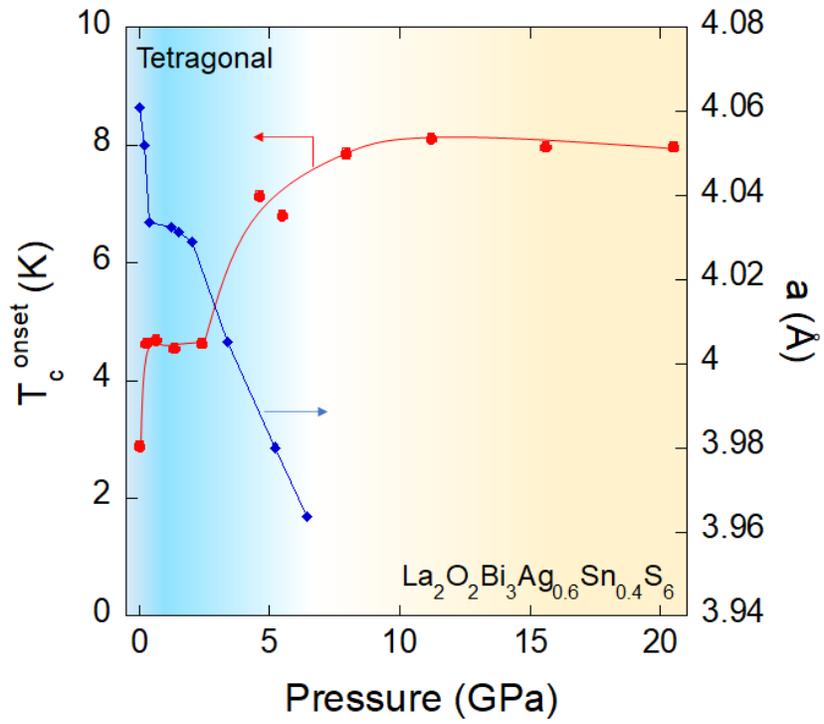

Figure 9. Pressure dependences of $T_c$ and lattice constant $a$ of $La_2O_2Bi_3Ag_{0.6}Sn_{0.4}S_6$.

**Supplemental Materials**

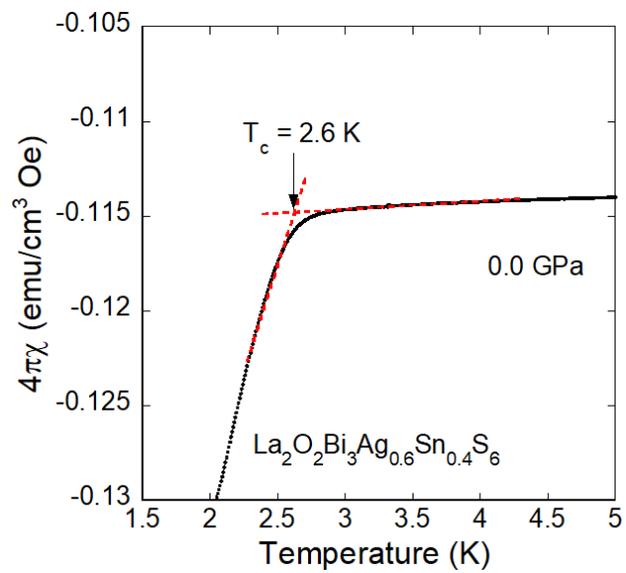

Fig. S1. Estimation of $T_c^{mag}$ for the magnetic susceptibility data at 0.0 GPa.

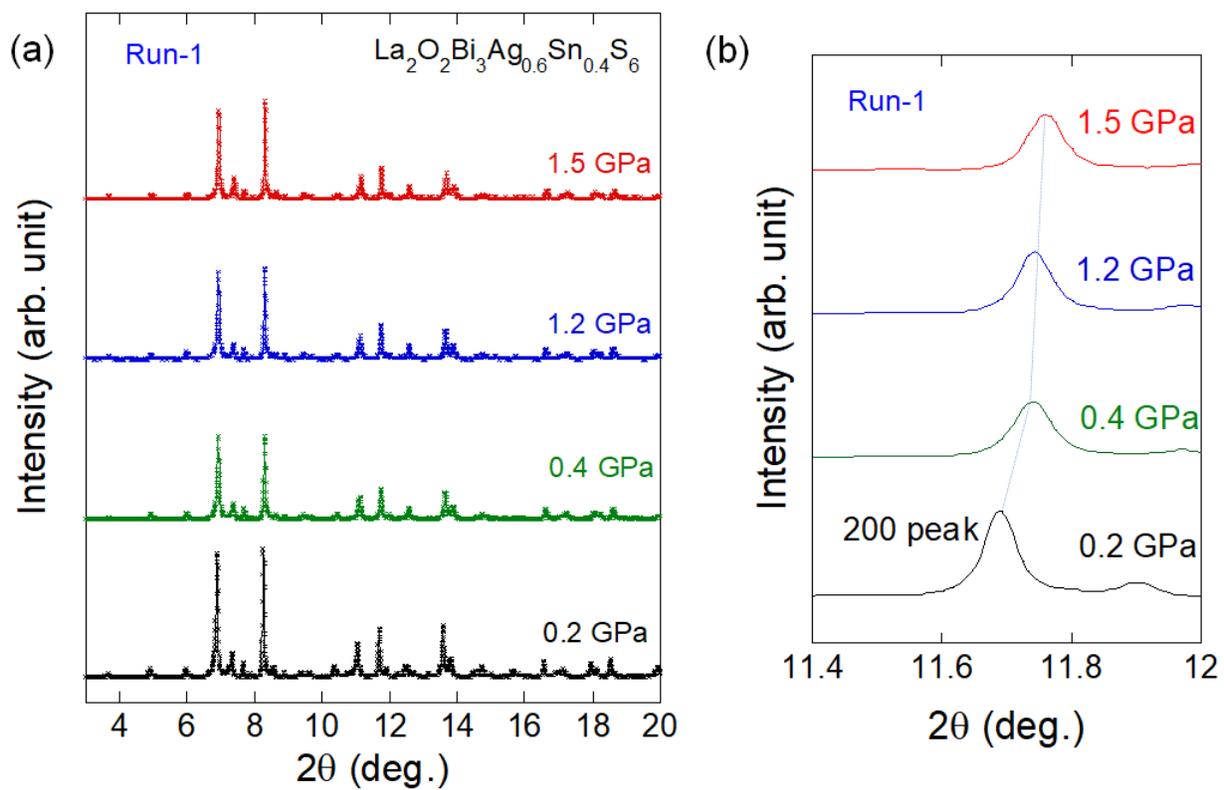

Fig. S2. (a) Powder SXRD patterns (Run-1) for $La_2O_2Bi_3Ag_{0.6}Sn_{0.4}S_6$. (b) Evolution of the 200 peak.

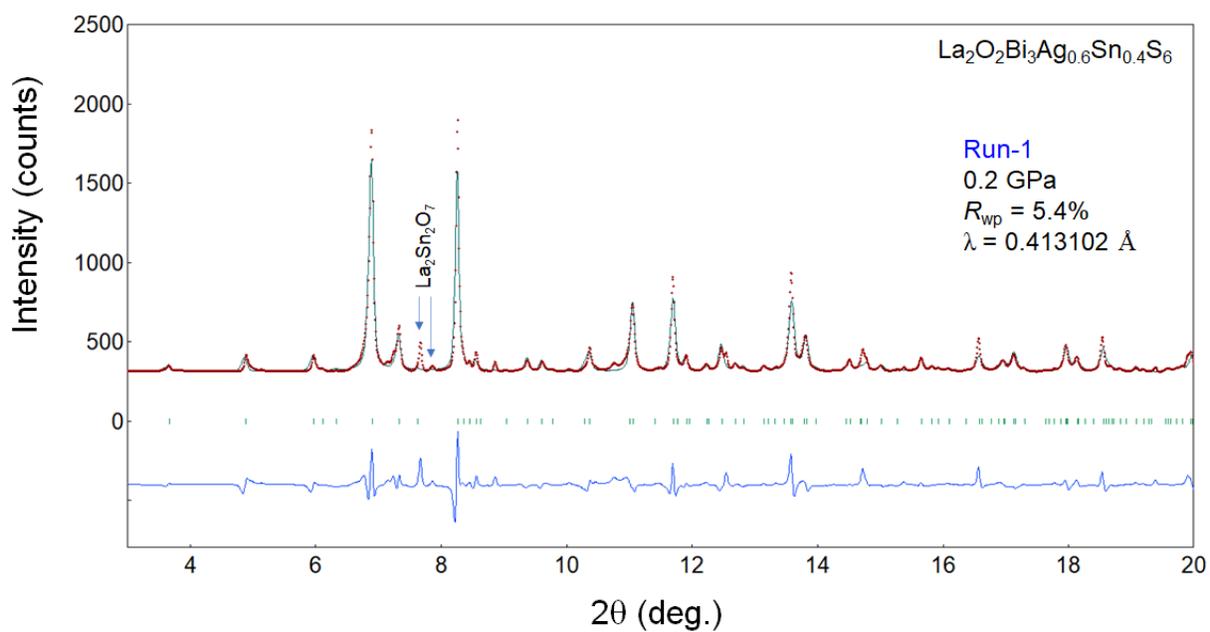

Fig. S3. Rietveld refinement for powder SXRD patterns (Run-1, $P$ = 0.2 GPa).

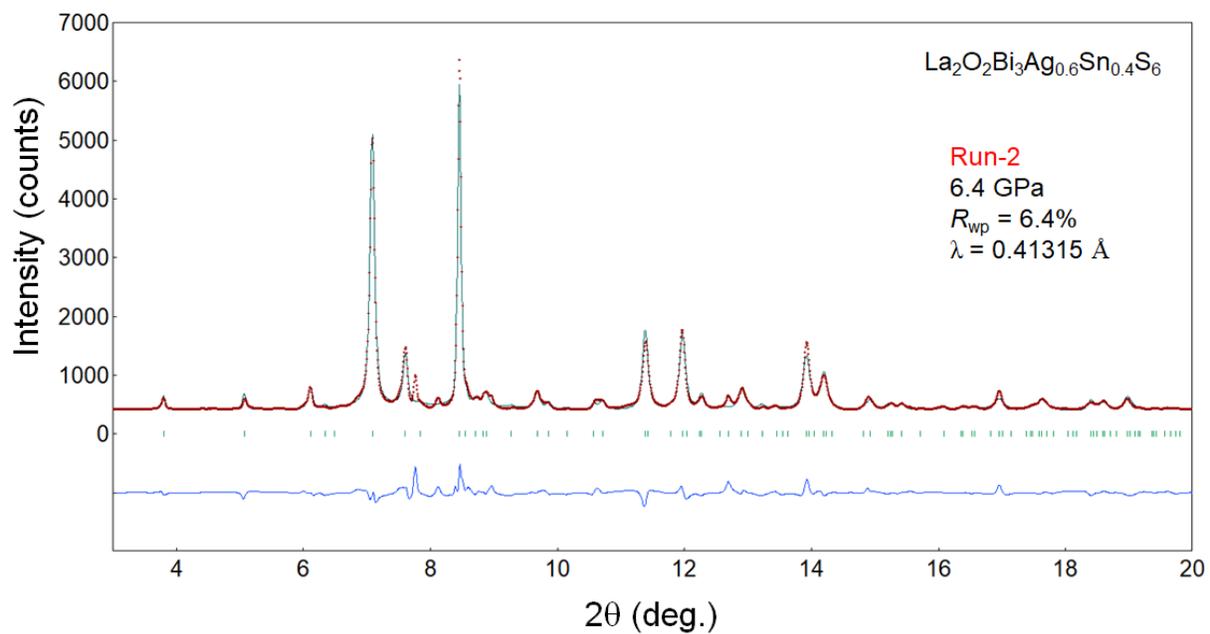

Fig. S4. Rietveld refinement for powder SXRD patterns (Run-2, $P$ = 6.4 GPa).